\documentclass[showpacs,longbibliography,superscriptaddress,twocolumn,prb]{revtex4-1}

\usepackage{pict2e}
\usepackage{bm}
\usepackage{graphicx}
\usepackage{amsmath}
\usepackage{amssymb}
\usepackage{xcolor}
\usepackage[percent]{overpic}
\usepackage{natbib}
\usepackage[pdftex,colorlinks]{hyperref}

\newcommand{\bra}[1]{\left\langle {#1} \right\vert}
\newcommand{\ket}[1]{\left\vert {#1} \right\rangle}
\newcommand{\bvec}[1]{{\mathbf{#1}}}

\begin{document}
\title{Tuning optical properties of Ge nanocrystals by Si shell}
\author{M.O.~Nestoklon} 
\author{A.N.~Poddubny}\affiliation{Ioffe Institute, 194021 St. Petersburg, Russia}
\author{P.~Voisin}\affiliation{Laboratoire de Photonique et Nanostructures, CNRS \\ and Universit{\'e} Paris-Saclay, Route de Nozay, 91460 Marcoussis, France}
\author{K.~Dohnalova}\affiliation{Van der Waals-Zeeman Institute, University of Amsterdam, Science Park 904, NL-1098 XH Amsterdam, The Netherlands}
\begin{abstract}
We present a theoretical study of Ge-core/Si-shell nanocrystals in a  wide bandgap matrix and compare the results with experimental data obtained from the samples prepared by co-sputtering. 
The empirical tight-binding technique  allows us to account for the electronic structure under  strain on the atomistic level. 
We find that a Si shell as thick as 1 monolayer is enough to reduce the radiative recombination rate as a result of valley $L-X$ cross-over. Thin Si shell leads to a dramatic reduction of the optical bandgap from visible to near-infrared range, which is promising for photovoltaics and photodetector applications.  

Our detailed analysis of the structure of the confined electron and hole states in real and reciprocal spaces indicates that the  type-II heterostructure is not yet achieved for Si shells with the thickness below 0.8 nm, despite some earlier theoretical predictions.  The energy levels of holes are affected by the Si shell stronger than the electron states, even though holes are completely confined to the Ge core. This occurs probably due to a strong influence of strain on the band offsets.

\end{abstract}
\maketitle

\section{Introduction}\label{sec:intro}
Ge nanostructures are of great interest for light emitter/detector applications in optoelectronics, photonics and photovoltaics.\cite{Suess13} The indirect bandgap $k$-vector selection rules are lifted in nanostructures by quantum confinement, promising greatly enhanced radiative rate and size-tunable bandgap.\cite{Niquet00, Bostedt04, Bulutay07, Sevik08, Ossicini10} Unfortunately, Ge nanocrystals (NCs) naturally oxidize upon exposure to air, leading to formation of O-related centers that act either as radiative centers with fast decaying blue-green emission \cite{Maeda95, Takeoka98, Niquet00, Chivas11} or as defect traps dramatically reducing emission (see e.g. remarks in Ref.~\onlinecite{Niquet00}). Despite these difficulties, Ge nanostructures remain of great interest, showing attractive properties like highly efficient space separated multiple exciton generation (MEG).\cite{Saeed14} Moreover, for integrated optoelectronics and photonics development, Ge remains to be cheaper and more CMOS compatible counterpart to Si than III-V and II-VI semiconductors.

A possible route to further advanced material properties engineering is alloying and straining of Ge by Si by developing various Si/Ge nano-heterostructures.\cite{Avezac12,Zhang12} Alloying Ge with Si allows one to tune bandstructure and bandgap into almost optimal range for photovoltaics employing carrier multiplication.\cite{Kolodinski95, Wolf98, Werner94} Tensile straining of Ge can lead to modified band structure towards direct bandgap,\cite{Walle89, Huo11, Jain12, Suess13} which is more pronounced compared to Si, due to a small energy difference between the direct $\Gamma$--$\Gamma$ and the indirect ($\Gamma$--$L$) transitions. Possible routes towards  permanent straining are core-shell type heterostructures, such as nanowires or nanocrystals. Si and Ge form type-II heterostructures and it is natural to expect that in the Si/Ge-based nanostructures the holes are localized in the Ge region and electrons in the Si region. However, there is a lattice mismatch of 4\% between the Si and Ge, and consequently, band alignment is modified by the strain.  Ge NCs capped with thin Si shell are also promising because Si acts as a protective coating, eliminating formation of the GeO oxide layer,\cite{Liu08} source of emissive defect centers \cite{Min96, Wu99} and/or emission quenching.\cite{Takeoka00} Type-II heterostructures are in general interesting for photodetectors and memory applications \cite{Liu08} due to lowered rates of optical transitions, as spatially separated electron and hole recombine less efficiently. Also, photon absorption process is typically controlled by the core-core or shell-shell transition, that is energetically shifted from the core-shell emission. Hence, such system also provides transparency for emission due to the lack of reabsorption, interesting for amplifiers and lasers. Core-shell scheme can also lead to  reduced blinking \cite{Galland12} and single exciton population inversion.\cite{Klimov07}

Quite recently, unusually bright and fast decaying ($\sim$1~ns) photoluminescence (PL) has been reported from co-sputtered Ge and Si-rich SiO$_2$.\cite{Saeed14_2} Three-component PL has been observed, strongly suggesting formation of type-II core-shell heterostructure, however, this suggestion was not confirmed. The properties of such core-shell nanocrystals were discussed in the literature mostly based on $k \cdot p$ approach that lacks sensitivity to valley-dependent phenomena,  crucial in understanding the electronic structure of the system. Also, not much emphasis has been put on complex strain in the realistic nanocrystals which could further obscure the details of their electronic structure.

The ways to prepare well-defined core-shell Si/Ge or Ge/Si NC systems are relatively scarce. In past, various techniques were used, such as mechanical alloying by ball milling,\cite{Wang08} wet chemical etching \cite{Malachias03} and molecular beam epitaxy,\cite{Alonso05, Valakh05, Demchenko07} wet chemical synthesis \cite{Yang99, Barry11}, synthesis from gas phase (silane and germane) \cite{Pi09, Erogbogbo11, Yasar13, Mehringer15} and co-sputtering.\cite{Takeoka00JL, Buljan10, Kolobov02} 
The segregation of Ge and Si is often the case for Ge-rich SiGe alloys, even though miscibility of Ge and Si is expected for whole compositional range. It has been shown both theoretically and experimentally that Si-capped Ge NCs have formation energies similar to those of Ge NCs with the same number of atoms, while Ge-capped Si NCs are less stable than Si NCs.\cite{Ramos05}  For samples with high Ge content prepared by radio frequency (RF) magnetron co-sputtering and annealed at high temperatures,  composite NCs consisting of a Ge NC core and amorphous SiGe shell have been reported from the Raman analysis.\cite{Kolobov02, Yang04}  
Currently, the best evidence for formation of Ge-core/Si-shell systems has been provided by M. Buljan using co-sputtering technique and X-ray analysis.\cite{Buljan10}

Available Density Functional Theory (DFT) calculations of small core-shell Si/Ge and Ge/Si NC systems by Ramos et al. \cite{Ramos05} and Oliveira et al. \cite{Oliveira12} predict type-II confinement with holes in Ge and electrons in Si. Si/Ge core-shell system is shown to offer high oscillator strength and long decay time, but large shift of absorption threshold with varying Ge shell thickness. Ge/Si core-shell, on the other hand shows long lifetimes with lower absorption shifts, making this material more suitable for e.g. photodetectors. For larger Si/Ge and Ge/Si core-shell NC systems, tight binding calculations are available by Neupane et al. \cite{Neupane11, Neupane15}, however, with limited scope: study  of carriers in Ge/Si core-shell system with thick fixed shell thickness of 5 nm \cite{Neupane11} and in large Ge/Si dome-shaped nanostructures with the base diameters in the range $5\div 45$~nm.\cite{Neupane15} In our work we address the missing category of larger Ge NC core with thin Si shell: system that can be relatively simply prepared by chemical synthesis \cite{Yang99} or sputtering.\cite{Buljan10} New tools have been developed to trace real- and $k$-space origin for each quantum-confined state. Using these tools we show that Ge optical properties can be extensively tuned by adding relatively thin Si shell.

\section{Tight-binding model}\label{sec:model}

\begin{figure}
\includegraphics[width=9cm]{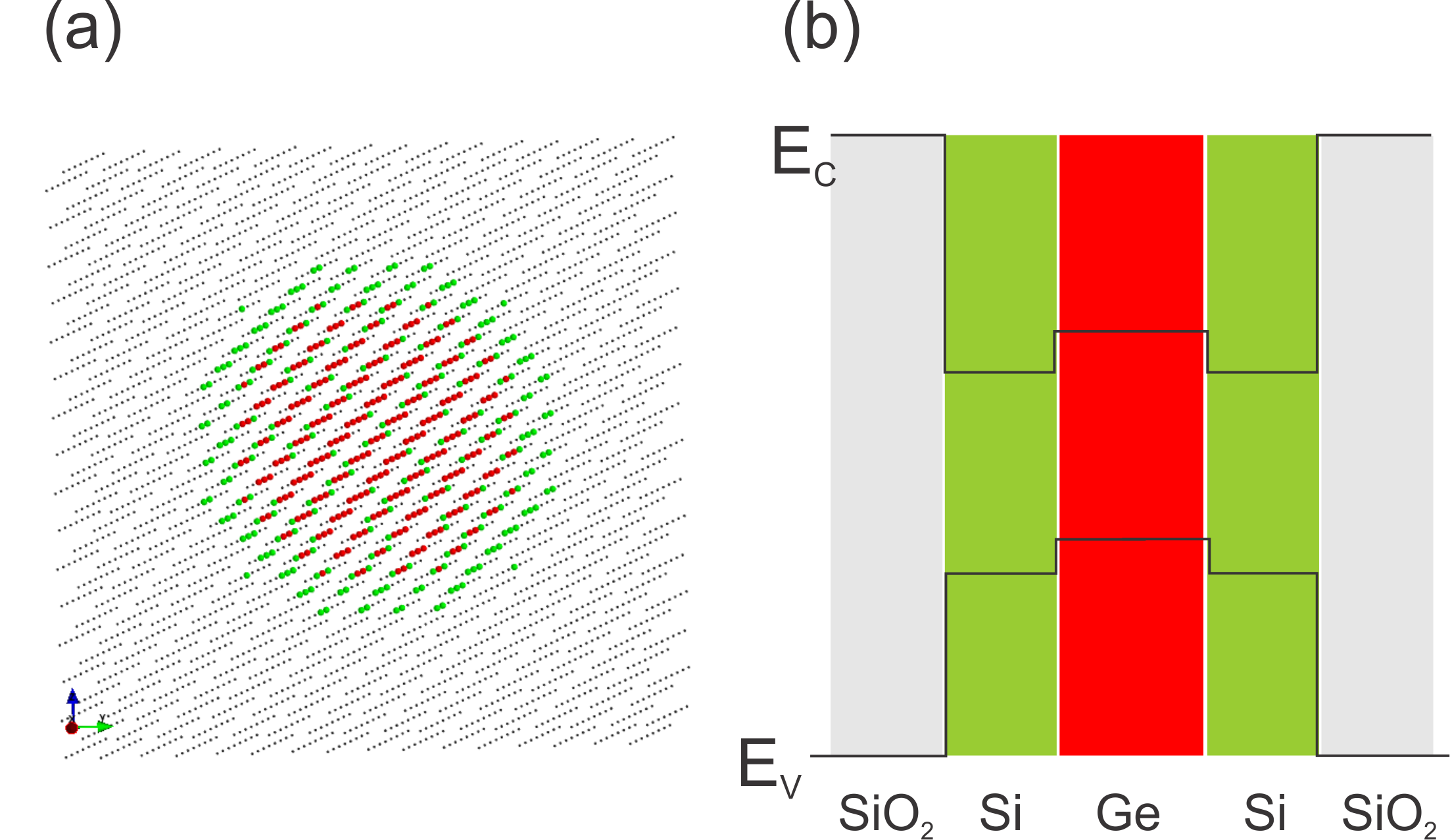}
\caption{(Color online) Schematic sketch of studied Ge/Si core/shell NCs system. (a) 3D plot of atomic arrangement in one of the real structures calculated. Ge atoms are shown by red dots, Si atoms are green and atoms which simulate SiO$_2$ matrix are gray.  (b) Sketch of nanocrystal band structure.}\label{fig:sample}
\end{figure}

\begin{figure}
\includegraphics[width=9cm]{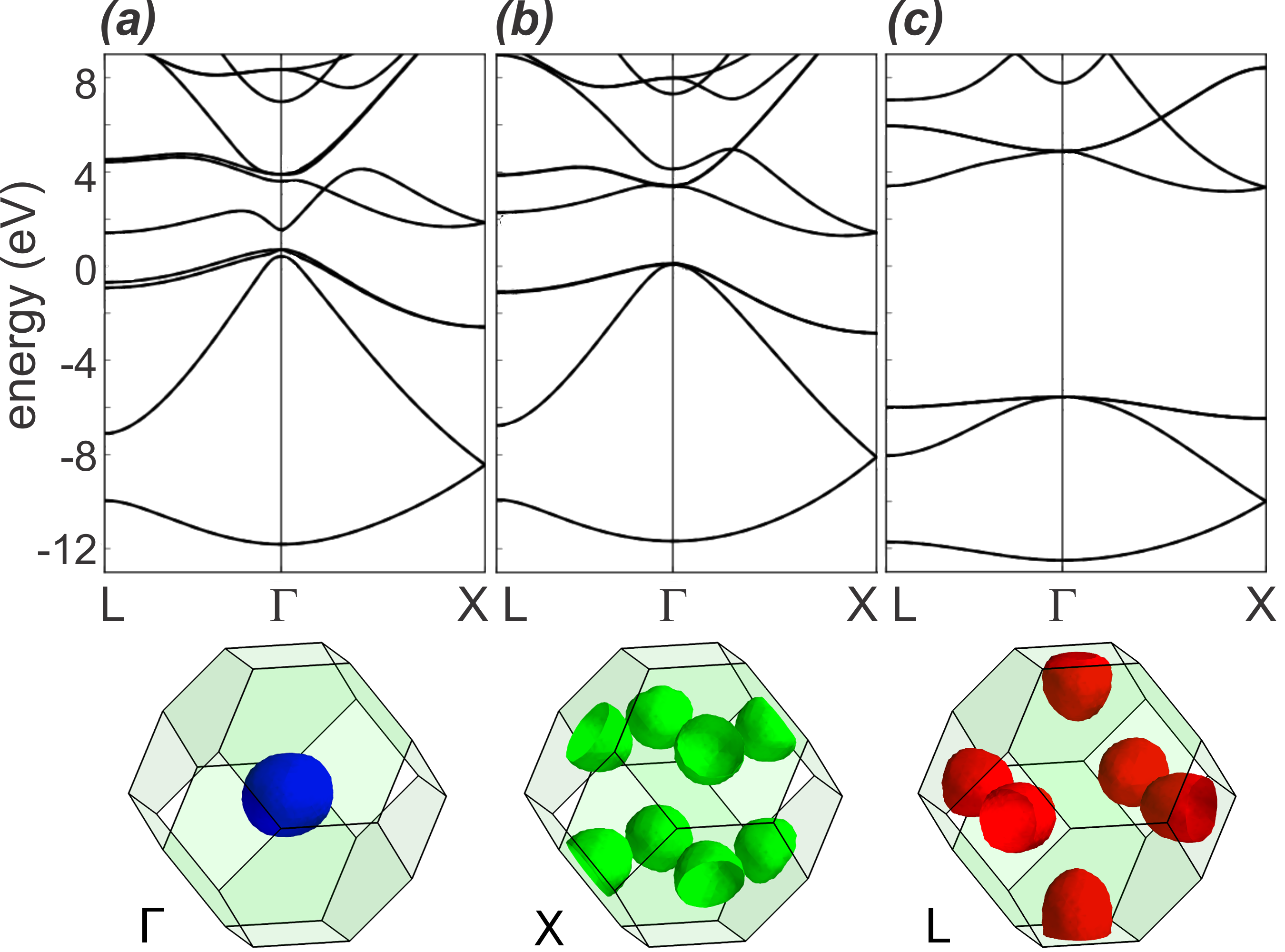}
\caption{(Color online) Tight-binding band structure of bulk (a) Ge, (b) Si and (c) SiO$_2$. (d) Brillouin zone with iso-surfaces  of $k$-space LDOS (see below) for typical electron states confined in $\Gamma$, $X$ and $L$ valleys for a real NC.}\label{fig:kspace}
\end{figure}

The structure under consideration is schematically shown in Fig.~\ref{fig:sample}(a). It consists of a Ge nanocrystalline core with the diameter $D_{\rm core}$ (red atoms), and crystalline Si shell with the diameter $D_{\rm shell}$ (green atoms) embedded in the amorphous SiO$_2$ matrix (gray atoms). The atomic coordinates between Si and SiO$_2$ were relaxed according to the standard valence force field (VFF) theory\cite{Keating66} in order to minimize the strain energy. This approach is reasonable for covalent semiconductors and provides atomistic strain with relatively high precision.\cite{Ruecker95} We model SiO$_2$ as a virtual crystal with zincblende lattice, with bandstructure near band gap similar to $\alpha$-quartz.   In order to describe nominally unstrained  core-shell Si/Ge NC, we set the VFF parameters to mimic soft material. The electronic spectrum has been obtained using the standard $sp^3d^5s^*$ empirical tight-binding (TB) technique\cite{Jancu98} and parametrization.\cite{Niquet09} Since there is no universally accepted TB parametrization of $\alpha$-SiO$_2$ in literature,\cite{Tersoff2010,watanabe2010} we have considered SiO$_{2}$ phenomenologically, with TB parameters manually chosen to reproduce the experimental band gap and offsets.\cite{Bulutay07} The valence and conduction band offsets between Si, Ge and SiO$_2$ lead to the type-II heterostructure system, the schematic band diagram is depicted in Fig.~\ref{fig:sample}(b). The tight-binding parameters used are given in supplementary material. 
The corresponding TB band structures of bulk  Ge, Si and SiO$_2$ are shown in Fig.~\ref{fig:kspace}(a-c), respectively. The bulk Si and Ge have quite different band structures with the conduction band minima located in the $X$ and $L$ valleys, respectively. The valence band extrema, however, are in the $\Gamma$ point for all considered materials. 

NC have discrete set of eigenstates with energy $E_i$ and wavefunction $\psi_i$.
The TB wavefunction of $i$-th state (we will omit it below when this does lead to confusion)
\begin{equation}
\psi_i=\sum\limits_{n,\alpha}C_{n\alpha}^{(i)}\Psi_{\alpha}(\bm r-\bm r_{n})
\end{equation}
 is determined by a set of complex coefficients $C^{(i)}_{n\alpha}$ defined on a lattice $\bm r_{n}$ in 3D space. The index $\alpha$ labels the orbital and spin degrees of freedom of the atomic-like wavefunctions $\Psi_{\alpha}(\bm r)$.
Tight-binding wave functions are hard to analyze without a postprocessing that extracts physically relevant quantities. An obvious example is the real space distribution of electron density: local density of states (LDOS) which we define as
\begin{equation}\label{eq:LDOS}
  n^{i}(\bm{r}) = \sum_{n,\alpha} |C_{n,\alpha}^{(i)}|^2 \frac{e^{-(\bm{r}-\bm{r}_n)^2/a^2}}{\left( \sqrt{\pi} a \right)^3}\,,
\end{equation}
where $C_{n\alpha}$ are the tight-binding coefficients at $n$-th atom, $a$ is Gaussian broadening which may be chosen to be of the order of interatomic distance and $\bm{r}_n$ is the position of $n$-th atom. Equation~\eqref{eq:LDOS} for LDOS allows one to trace the real space distribution of the electron density for the particular state. However, LDOS only is not enough to visualize all the features of electron states. In order to trace the valley structure of the states, below we define k-space LDOS. For this, we perform a Fourier transform of the  TB wavefunction of confined carriers as explained below. For a given lattice $\bm r_{n}$ one can  formally define a lattice in the reciprocal space.\cite{ITCB} For the finite system we consider a large supercell and apply periodic boundary conditions. The supercell is defined as a parallelogram cut from ideal zincblende lattice by vectors $\bvec{A}_i,\; i=1..3$. This parallelogram defines a lattice in the reciprocal space with the basis vector
\begin{equation}
   \bvec{ B}_1 = 2\pi \frac{\bvec{A}_2 \times \bvec{A}_3}
    {\bvec{A_1}\cdot [\bvec{A}_2 \times \bvec{A}_3]}\:,
\end{equation}
and $\bvec{B}_{2}$, $\bvec{B}_{3}$ are obtained by the cyclic permutations. 
\setlength{\unitlength}{0.1cm}
\begin{figure}
\begin{picture}(84,30)
  {\color{cmyk:black,1;white,10} \polygon*(0,2.5)(25,5)(28.5,26.0)(3.5,23.5)}%
  \put(0,2.5){\linethickness{0.2mm}\vector(5,0.5){5}}
  \put(0,2.5){\linethickness{0.2mm}\vector(0.5,3.0){0.5}}
  \multiput(0,2.5)(5,0.5){6}{\multiput(0,0)(0.5,3.0){8}{\circle*{1}}}%
  \put(0,2.5){\linethickness{0.15mm}\color{blue}\vector(25,2.5){25}}
  \put(0,2.5){\linethickness{0.15mm}\color{blue}\vector(3.5,21.0){3.5}}
  \multiput(4.5,-0.5)(5,0.5){6}{\color{cmyk:black,1;white,1}\circle*{1}}
  \multiput(30,5.5)(0.5,3){8}{\color{cmyk:black,1;white,1}\circle*{1}}
  \multiput(4.0,26.5)(5,0.5){6}{\color{cmyk:black,1;white,1}\circle*{1}}
  \multiput(4.5,29.5)(5,0.5){1}{\color{cmyk:black,1;white,1}\circle*{1}}
  \put(2.5,0.5){\makebox[0pt][c]{$\bm{a}_1$}}%
  \put(-1.75,3.25){\makebox[0pt][c]{$\bm{a}_2$}}%
  \put(23.5,2.5){\makebox[0pt][c]{$\bm{A}_1$}}%
  \put(1.05,21.0){\makebox[0pt][c]{$\bm{A}_2$}}%
  {\color{cmyk:black,1;white,10} \polygon*(50,6)(85,4.0)(78,25)(43,27)}%
  \put(50,6){\linethickness{0.3mm}\vector(35,-2.0){35}}
  \put(50,6){\linethickness{0.3mm}\vector(-7,21){7}}
  \put(50,6){\linethickness{0.15mm}\color{blue}\vector(7,-0.4){7}}
  \put(50,6){\linethickness{0.15mm}\color{blue}\vector(-1,3.0){1}}
  \multiput(50,6)(7,-0.4){6}{\multiput(0,0)(-1,3.0){8}{\circle*{1}}}%
  \multiput(44,3.4)(7,-0.4){6}{\color{cmyk:black,1;white,1}\circle*{1}}%
  \multiput(44,3.4)(-1,3){3}{\color{cmyk:black,1;white,1}\circle*{1}}%
  \multiput(49,29.6)(7,-0.4){5}{\color{cmyk:black,1;white,1}\circle*{1}}%
  \put(81.0,1.5){\makebox[0pt][c]{$\bm{b}_1$}}%
  \put(42.0,22){\makebox[0pt][c]{$\bm{b}_2$}}%
  \put(53.5,2){\makebox[0pt][c]{$\bm{B}_1$}}%
  \put(47.5,6){\makebox[0pt][c]{$\bm{B}_2$}}%
\end{picture}
\caption{Illustration of atom real space ($\bm{a}_i$, $\bm{A}_i$) and reciprocal ($\bm{b}_i$, $\bm{B}_i$) space basis vectors of lattice ($\bm{a}_i$, $\bm{b}_i$) and supercell ($\bm{A}_i$, $\bm{B}_i$).
}\label{fig:r_n_k}
\end{figure}
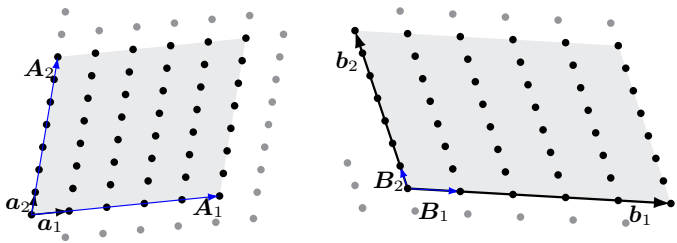

Positions of cations in the ideal lattice are defined as
\begin{equation}
    \bvec{r}_{n_1n_2n_3} = \bvec{a}_1 n_1 + \bvec{a}_2 n_2 + \bvec{a}_3 n_3\:,
\end{equation}
where $\bvec{a}_i$ are the standard basis vectors of face-centered cubic lattice and anions are shifted by the vector $a_0/4(1,1,1)$. 
The vectors $\bvec{a}_i$ correspond to the reciprocal space vectors
\begin{equation}
   \bvec{b}_1 = 2\pi \frac{\bvec{a}_2 \times \bvec{a}_3} {\bvec{a_1}\cdot [\bvec{a}_2 \times \bvec{a}_3]}\:, \text{and c.p.}
\end{equation}
The vectors $\bvec{b}$  form the basis of another lattice in the reciprocal space associated with the crystal lattice. The  
first Brillouin zone of the lattice is defined by the supercell vectors $\bm{B}_i$ by considering vectors $\bvec{k}_{n_1,n_2,n_3}=\bvec{B}_1 n_1 + \bvec{B}_2 n_2 + \bvec{B}_3 n_3$ equal if and only if they are differ by a linear combination of $\bvec{b}_i$. We will call this set of vectors ``k-space mesh''
\begin{multline}
  \mathcal{K}_{\text{BZ}} = \big\{ \bvec{k}_{n_1,n_2,n_3} \big\vert \bvec{k}_{n_1,n_2,n_3}-\bvec{k}_{n_1',n_2',n_3'} \neq  m_1\bvec{b_1} \\ + m_2\bvec{b_2} + m_3\bvec{b_3} \big\}
\end{multline}
Simple linear algebra shows that the number of vectors $\bvec{k}_{n_1,n_2,n_3}$ in the first Brillouin zone is exactly equal to the number of prime cells in a supercell bounded by the vectors $\bvec{A}_{i}$.
Definition of the vectors $\bvec{a}_i$, $\bvec{A}_i$, $\bvec{b}_i$ and $\bvec{B}_i$ is illustrated in Fig.~\ref{fig:r_n_k}. 
With these definitions we introduce a discrete Fourier transform by
\begin{equation}\label{eq:DFT}
    C^{\bvec{k}}_{l\tau\alpha} = \sum_{n\in\text{sublattice $\tau$}}
    C_{n\alpha} e^{i\bvec{r}_n\bvec{k}_l}, \;\; \bvec{k}_l \in \mathcal{K}_{\text{BZ}}\,,
\end{equation}
that gives us the same number of coefficients as we had in the TB wavefunction in real space. The sum in Eq.~\eqref{eq:DFT} runs through atoms in two sublattices separately. This formal definition satisfies the obvious properties one could expect from a function in reciprocal space: the norm is conserved, the orthogonality of the states is preserved. If one considers the state with well-defined wave vector $\bvec{k}_{0}$, the corresponding Fourier transform $C^{\bvec{k}}_{l\tau\alpha}$ has a sharp maximum for $\bvec{k}_l \sim \bvec{k}_{0}$. This allows us to attribute the confined states to particular valleys.

Similar to LDOS in real space Eq.~\eqref{eq:LDOS} we may define LDOS in reciprocal space, which we will call k-space LDOS or kLDOS below, as 
\begin{equation}\label{eq:kDOS}
  n^i(\bm{k}) = \sum_{l\tau\alpha} |C^{\bvec{k}(i)}_{l\tau\alpha}|^2 \frac{e^{-(\bm{k}-\bm{k}_l)^2/q^2}}{\left( \sqrt{\pi} q \right)^3}\,,
\end{equation}
where the $k$-space Gaussian smoothing parameter $q$ is conveniently chosen of the order of $|\bvec{B}_i|$.
To illustrate this procedure, Fig.~\ref{fig:kspace}(d) presents the isodensity surface plots of $k$-space LDOS calculated using Eq.~\eqref{eq:kDOS} for few particular states which show definite $\Gamma$, $L$ and $X$ valley character for a realistic calculations of NCs, see below. It can be seen that the kLDOS indeed clearly shows the valley structure.

Importantly, this approach can  be formally used for strained structures as well, although care is needed. When the strain is applied, atomic positions start to deviate from ideal crystal lattice and a proper generalization of the above scheme must be used. The approach outlined above is closely related with the unfolding scheme used in Ref.~\onlinecite{Hapala13} and up to some generalizations and approximations it is equivalent to the unfolding scheme proposed in Refs.~\onlinecite{Boykin05, Boykin07, Zunger10, Zunger12}, but the detailed comparison is out of the scope of the current paper.

\section{Results for varying core and shell thicknesses}\label{sec:results}
Now we proceed to the detailed analysis of influence of the crystalline Si shell on electronic structure and density of states (DOS) re-distribution with respect to the Ge NC without shell. 
We have simulated many different systems that can be separated into two major groups: (a) systems with fixed Ge NC core diameter to 2.9 nm and variable shell of the thickness 0, 0.12, 0.25, 0.54, 0.78, 0.93 and 1.06 nm; (b) set of five different Ge NC core sizes of 1.50, 1.75, 2.25, 2.75 and 3.50~ML (monolayers), which corresponds to 1.7, 2.0, 2.5, 3.1 and 3.9 nm core diameter, each studied for a set of Si shell thicknesses of 0, 0.12, 0.54 and 0.80~nm. More details on the second set can be found in the Supplementary Information. We model SiO$_2$ as a soft material, as a result the strain at the NC surface is relaxed. However, the strain between the Ge core and Si shell (which roughly results in compression of Ge core and stretching of Si shell) is fully included in the simulations and contributes to modifications in the band structure.

\begin{figure}
\includegraphics[width=\linewidth]{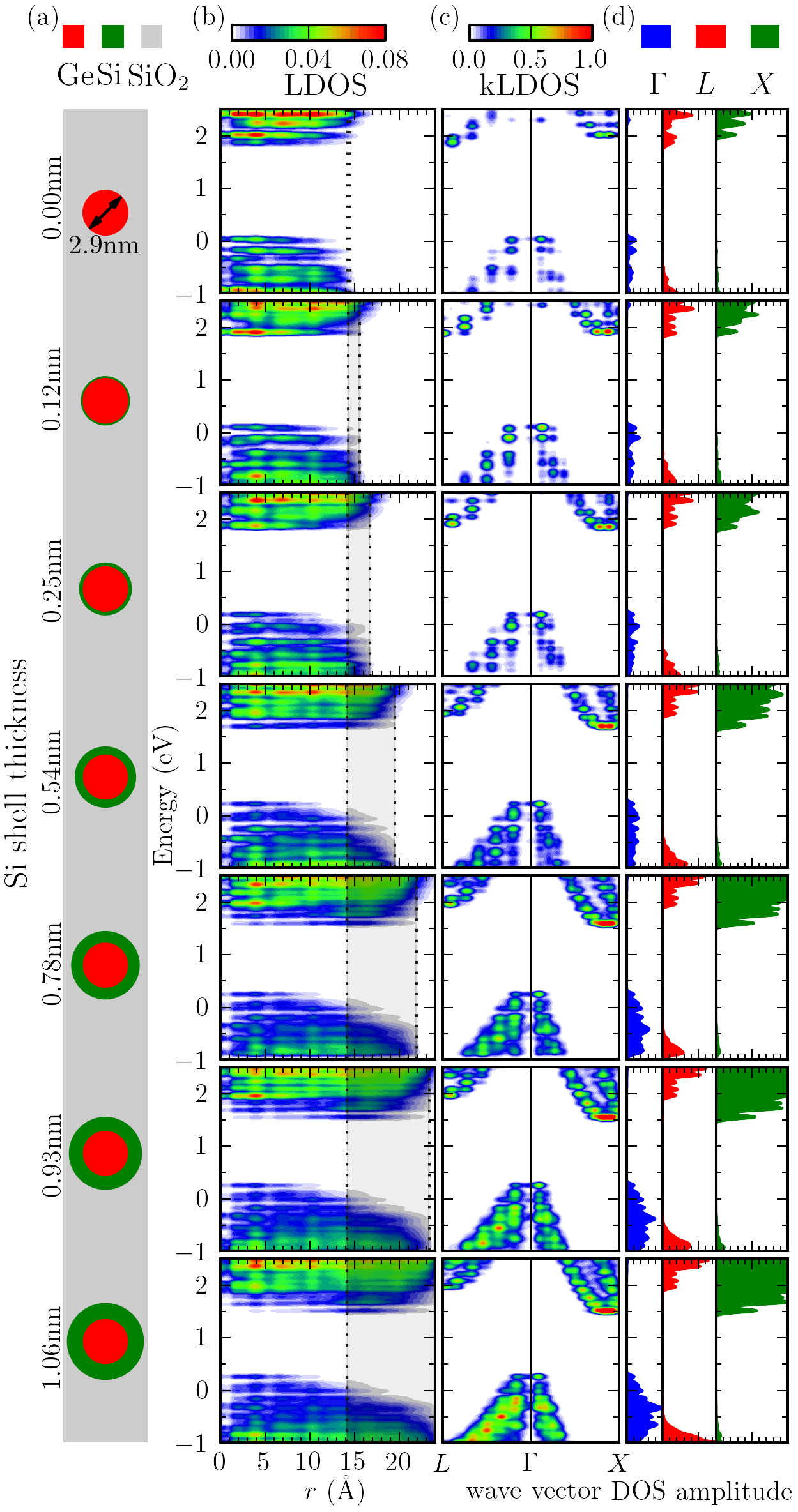}
\caption{(Color online) (a) Schematic sketch of the studied Ge/Si core/shell NC systems with the fixed core diameter of 2.9~nm and variable shell thickness between 0.00 and 1.30 nm.  (b) real space energy-resolved LDOS, Eq.~ \eqref{eq:rEDOS}. The grey area shows the position of the Si shell. (c) reciprocal space energy-resolved LDOS, Eq.~\eqref{eq:kEDOS} with $\bvec{k}$ varying along the $L$--$\Gamma$--$X$ path. (d)  valley-resolved DOS. Three panels show DOS in $\Gamma$ (blue), L (red) or X (green) valley. 
}
\label{fig:shellbandstructure}
\end{figure}

First we will study fixed Ge NC core size of 3.1 nm with the Si shell varied between 0 and 1.30 nm (see Fig.~\ref{fig:shellbandstructure}(a)). Figure~\ref{fig:shellbandstructure}(b) presents the energy-resolved LDOS for hole and electron states in the radial direction, 
\begin{equation}\label{eq:rEDOS}
  n(r,E) = \frac1{r^2} \sum_i \sum_{n,\alpha} \left| C_{n\alpha}^{(i)} \right|^2 e^{-\frac{(r-r_n)^2}{a_r^2}}e^{-\frac{E-E_i}{a_{E}^{2}}}.
\end{equation}
where the index $n$ enumerates the atoms with the distance to the QD center $r_n$, $i$ enumerates different solutions with the energies $E_i$ and two broadening parameters $a_r = 1$~\AA{} and $a_E = 50$~meV. An extra factor $1/r^2$  is added to \eqref{eq:rEDOS} to produce the values analogous to the regular DOS that satisfy the condition $n(E)\sim \int n(r,E)d^3\bvec{r}$. Color indicates high (red) to low (blue) DOS and the vertical gray lines denote edges of the core and the shell, respectively, with the shell area shown by the  gray rectangle. As  expected, for thicker Si shell, electrons become  localized in the shell, however, due to small energy offset between Si and Ge conduction band minima, the localization degree is relatively weak. 

It is instructive to analyze the microscopic structure of the confined states also in the wave-vector space. This can be shown by examining the energy-resolved k-space LDOS (the term ``fuzzy band structure'' has been used in Ref.~\onlinecite{Hapala13}) which is calculated as:
\begin{equation}\label{eq:kEDOS}
  n(\bvec{k},E) = \sum_i \sum_{l,\tau,\alpha} \left| C^{\bvec{k} (i)}_{l\tau\alpha} \right|^2 e^{-\frac{(\bvec{k}-\bvec{k}_l)^2}{a_k^2}} e^{-\frac{(E-E_i)^2}{a_E^2}}\:,
\end{equation}
where $a_k = 0.6$~\AA$^{-1}$ and $a_E = 50$~meV. 
 Then we fold k-space onto irreducible part of the Brilllouin zone using its point symmetry and take the 2D $xy$ slice of the function $n(\bvec{k},E)$ with $\bvec{k}$ running along the $L$--$\Gamma$--$X$ path in reciprocal space as ``$x$'' axis and energy as ``$y$'' component. Note that we do not consider integration over angle here, unlike in \eqref{eq:rEDOS}: the real space distribution of LDOS is more or less isotropic while resiprocal space LDOS is noticably localized near valley minima (see Fig.~\ref{fig:kspace}(d)). The results of calculation are presented in Fig.~\ref{fig:shellbandstructure}(c). Colors indicate high (red) to low (blue) DOS.  Notably, even in this relatively small Ge NCs, the band structure retains the features of the bulk. When the Si shell thickness increases, the energy of the ``$L$-branch'' slightly increases and, simultaneously, the ``$X$-branch'' energy decreases, see Figs.~\ref{fig:shellbandstructure}(c).  

In Fig.~\ref{fig:shellbandstructure}(d) we evaluate the valley composition of the confined states quantitatively by integrating the density in the reciprocal space in the regions near $X$, $\Gamma$ and $L$ points. The result is a DOS in three valleys separately. Figure~\ref{fig:shellbandstructure}(d) shows, that the valence band states are constructed mostly from the states near $\Gamma$ point. On the other hand, the structure of the confined conduction band states changes from $L$-valley states to an $X$-valley states (valley cross-over) when the Si shell is added and its thickness increases. This effect is more pronounced in smaller Ge NCs and  has a threshold behavior.

\begin{figure}
\includegraphics[width=9cm]{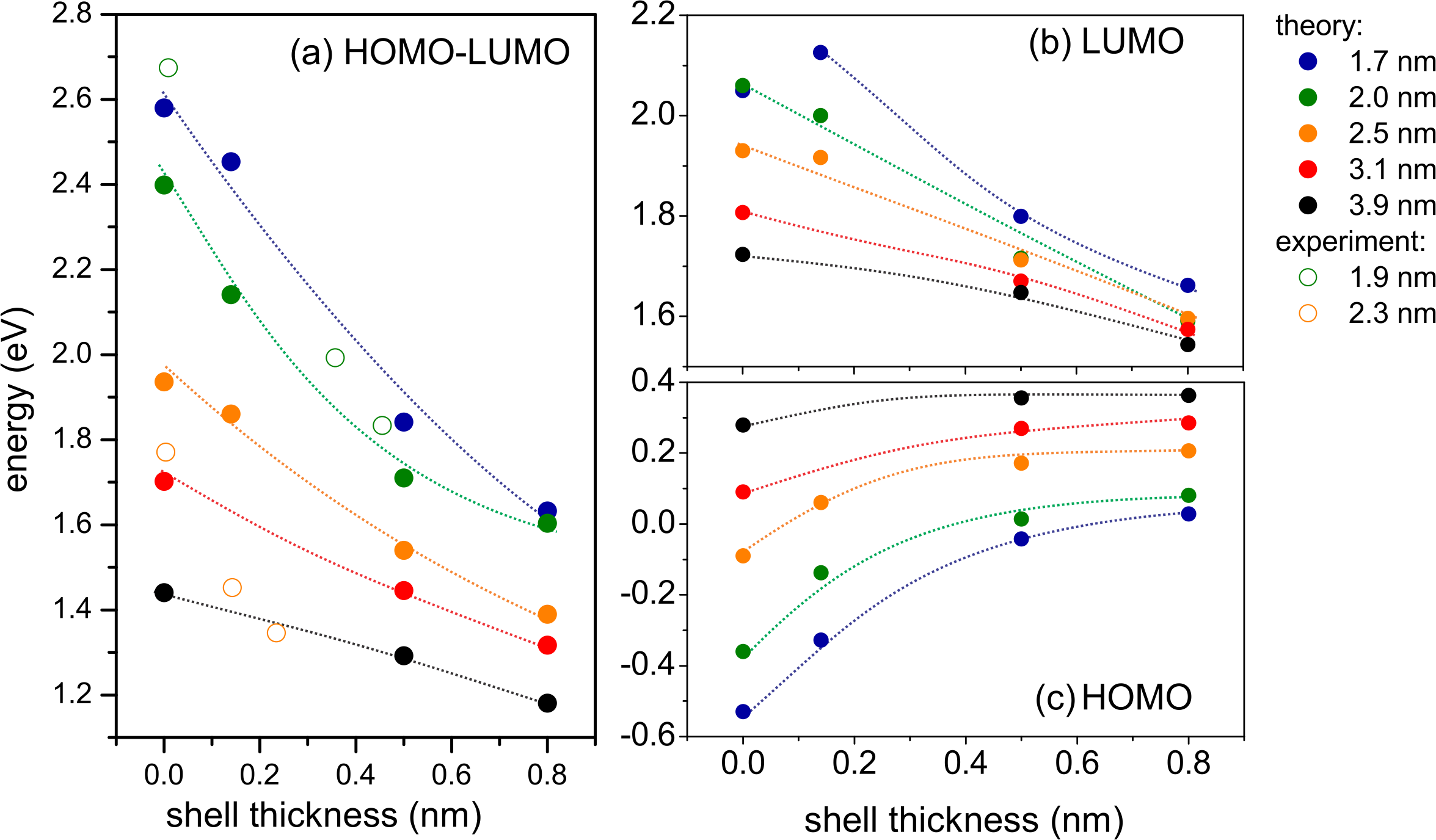}
\caption{(Color online) (a) Band gap energies and (b) electron (LUMO) and (c) hole (HOMO) energies as a function of the NC core size and shell thickness.  Dotted lines serve as the quide to the eye. Empty circle data are experimental data obtained by M. Buljan \cite{Buljan10} for Ge-Si core-shell structures (with not completely enclosed shells).}\label{fig:gap}
\end{figure}

It is interesting that  the Si shell leads not only to the valley cross-over for the conduction band from $L$ to $X$, but also to a dramatic reduction in the band gap energy of the whole system. Figure~\ref{fig:gap}(a) demonstrates that the NC band gap strongly decreases for larger core diameters as expected from the size quantization effect, but it is also reduced to a similar degree by adding the Si shell only. The shell-induced reduction is more pronounced for smaller NCs. Interestingly, even though holes are confined in the Ge core (Fig.~\ref{fig:shellbandstructure}(b)), their energy (HOMO) is more sensitive to the shell presence than that for the electronic states (LUMO), which are delocalized over the Ge core and Si shell (more localized in Si for thicker shells), see Fig.~\ref{fig:shellbandstructure}(b). This is related to larger band offset for the valence band than for conduction band. For thicker Si shell, the spectral tunability by Ge NC core size is reduced when compared to the systems with thinner shell. We would like to note here, that  the native bandgap for uncapped Ge NCs in SiO$_2$ here is reduced with respect to the hydrogen-capped Ge NC in vacuum, as a result of lower potential barrier at the Si/SiO$_2$ boundary. Optical bandgap energies obtained in our model are compared in Fig.~\ref{fig:gap}(a) with the experimental data obtained in the group of M. Buljan \cite{Buljan10} from the analysis of absorption and structural properties of Ge/Si core-shell samples prepared by co-sputtering in Al$_2$O$_3$ matrix. The calculation in Fig.~\ref{fig:gap} has been performed for a cubic supercell with the edge length few ML larger than NC diameter (large enough to neglect tunneling through SiO$_2$). Theoretical and experimental data appear to agree quantitatively and qualitatively  for smaller Ge NCs. For larger NCs the measured band gap  is smaller  than that predicted here. However, we note  that structures prepared by M. Buljan have not fully closed Si shell, as revealed by their in-depth structural analysis by the small-angle X-ray scattering (SAXS).\cite{Buljan10}

Now we proceed to the analysis of the radiative recombination rates depending on the thickness of the Si shell for the 3.1 nm Ge NCs. The phonon-less optical transition rate between the electron states $|e\rangle$ and $|h\rangle$ in the conduction and valence bands is given by \cite{Delerue04}
\begin{equation}
    \frac1{\tau} = \frac{4e^{2} \sqrt{\varepsilon_{\rm out}}(E_{e}-E_{h})}{3\hbar^{2} c^{3}}\mathcal F^{2} \left|\bra{e} v \ket{h}\right|^2\:,
    \label{eq:rate}
\end{equation}
where $\mathcal F=3\varepsilon_{\rm out}/(2\varepsilon_{\rm out}+\varepsilon_{\rm in})$ is the local field factor, $E_{e}$ and $E_{h}$ are the energies of the corresponding states. To account for the possible degeneracy of the states the rate Eq.~\eqref{eq:rate} is to be summed over the final states $|h\rangle$ with the same energy and averaged over the initial states $|e\rangle$. The velocity operator $\bvec{v}$ can be defined as a commutator of the tight-binding Hamiltonian with the coordinate operator, $\bvec{v} = i[H,\bvec{r}]/\hbar\:.$ We use the diagonal approximation $\bvec{r} = \delta_{nn'}\delta_{\alpha\beta} \bvec{r}_{n}$ for the coordinate operator.\cite{RamMohan93, JETP, Sandu05} This corresponds to neglecting the intraatomic matrix elements of the velocity operator for the optical transitions, which is reasonable for covalent semiconductors.\cite{Cruz99}

\begin{figure}
\includegraphics[width=\linewidth]{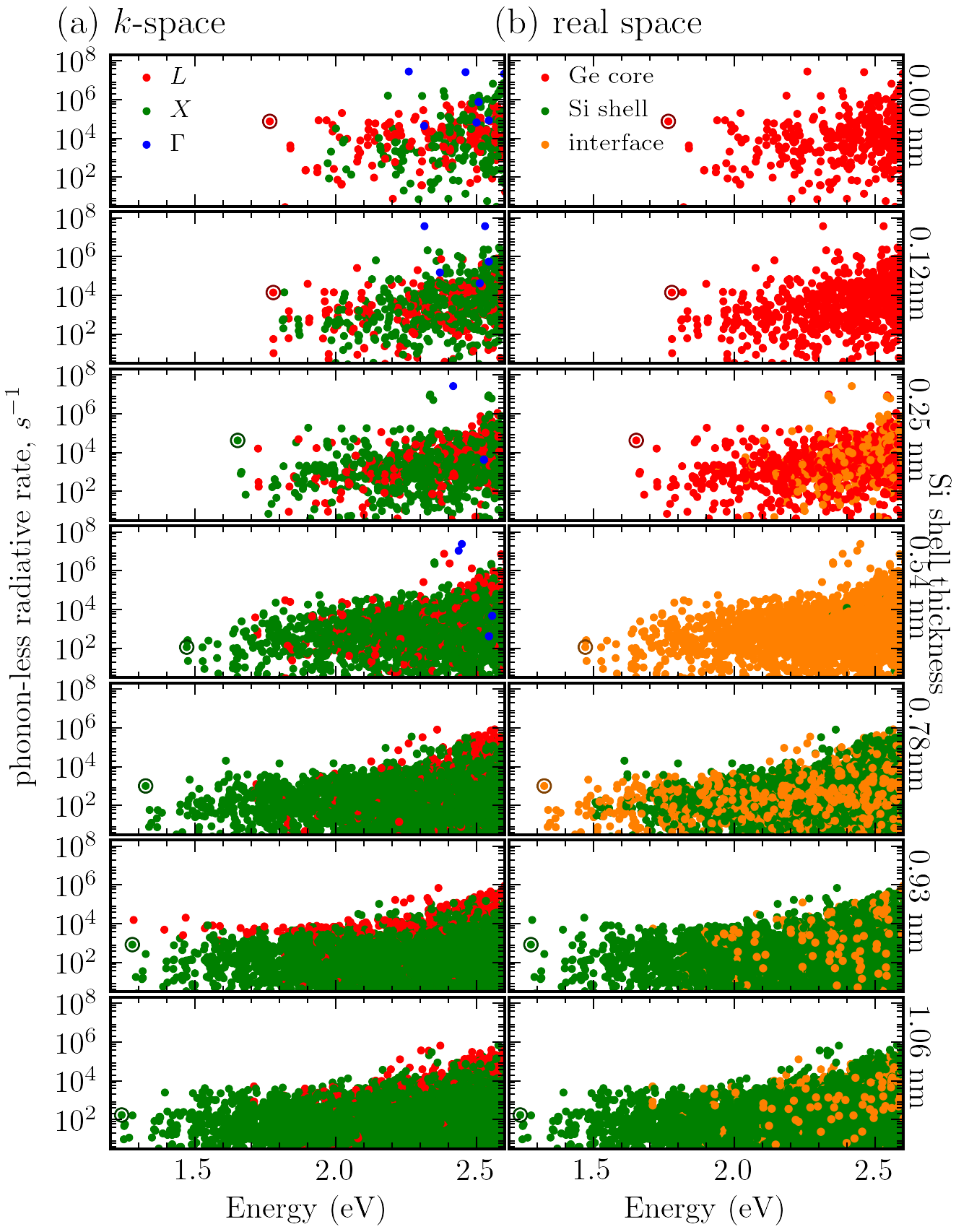}
\caption{(Color online) Radiative phonon-less transition rates resolved in color in k-space (a) and real-space (b). Ground state transition is highlighted by circle around it.}\label{fig:shellrates}
\end{figure}

The calculated phonon-less radiative transition rates are presented in Fig.~\ref{fig:shellrates}. Each transition is shown by a colored dot. To highlight the  valley structure of the electron states, the color in Fig.~\ref{fig:shellrates}(a) 
indicates the valley attribution of  the state, i.e. the blue color corresponds to the  electron density over 40\% in the $\Gamma$ point, green to the $X$ point and red to the $L$ point. The figure clearly demonstrates the decrease of the band gap for thicker Si shell, as presented previously in Figs.~\ref{fig:shellbandstructure} and \ref{fig:gap}. Also, the valley cross-over is clearly visible, as the lowest energy transitions change from L-valley (red) to X-valley (green) for thicker shell. Furthermore, addition of shell also leads to dramatic reduction in the rates, which appears to be clearly related to the modification of the $k$-space structure of the electron states.

It is interesting to plot also the real-space-resolved transition rates (Fig.~\ref{fig:shellrates}(b)) with the color signifying more than 40\% localization of the electron density component in Ge core (red), Si shell (green) or delocalized equally in between (orange). This figure shows that even though the localization of electron in the shell in Fig.~\ref{fig:shellbandstructure}(b) was not clearly pronounced, the rates are already reduced  for 0.5~nm-thick Si shell. As such, the rates suppression slightly precedes the over 40\% localization of the electron in the shell, which occurs only for shells thicker than 0.8 nm. This means that the valley cross-over has already a strong influence on the rate reduction beyond the ``standard" spatial delocalization role found in typical II-type heterostructures.

\section{Conclusions}\label{sec:concl}

In our simulation, we present a detailed analysis of Ge/Si core-shell NC systems that can be prepared by wet chemical synthesis \cite{Yang99} or co-sputtering.\cite{Buljan10} New tools have been developed to trace real- and reciprocal-space origin for every transition.

Since the Si shell leads to compressive strain of Ge core, the direct bandgap cannot be induced in such structures. 
It has been previously shown by the DFT ab-initio simulations \cite{Ramos05, Oliveira12} that the Si/Ge and Ge/Si core-shell NC systems are type-II heterostructures with electrons located in Si and holes in Ge. 
In our simulations for larger systems, we found that the two phenomena should be distinguished, namely the localization of an electron in the Si shell in the real space and the $L$-to-$X$ valley crossover  of the conduction band minimum in the reciprocal space. Localization in the shell is not quite distinct and can be seen only for lowest conduction band states when the shell thickness exceeds approximately 0.7 nm. On the other hand, the valley crossover is already seen for diatomic Si shell. The valence band maximum remains in the $\Gamma$  point for all simulated Si shell thicknesses. 
Interestingly, even though the holes reside in the Ge core, their energy (HOMO) is influenced by the Si shell more than that of electrons (LUMO). This effect is more pronounced in smaller NCs. 

The crossover and associated real space localization of the wavefunction leads to a dramatic reduction in the phonon-less radiative recombination rates.  Reduced rates are of interest for photodetector applications.
Improved emission yield might be expected for Si-capped NCs, since reduced recombination rate also leads to reduced Auger recombination.

The emission energy of Ge QDs can be efficiently tuned by adding Si shell of variable thicknesses. For fixed NC core size, strong reduction in band gap energy was observed with thicker Si shell, from visible range towards near-infrared range, interesting for photodetectors and photovoltaics. The optical bandgap energy and shell-dependence are in a good quantitative and qualitative agreement with experimental data obtained by M. Buljan \cite{Buljan10} from samples prepared using co-sputtering technique.
For thicker Si shell, the core-size-tunability of the optical bangap is drastically reduced. This is mainly related to strain induced by the Si shell, which modifies the energy offsets of valence and conduction bands differently. This means that the otherwise steep tunability of Ge NCs can be limited by the Si shell, allowing for achieving narrower spectral range for broader-size ensemble of NCs. 

In addition to the calculation presented here, the $L$--$X$ crossover may lead to significant change of phonon-assisted transitions. In order to develop  the detailed theory of optical properties of Ge/Se core-shell systems one also needs an accurate consideration of non-radiative recombination at defects in SiO2 and the renormalization of exciton energy by dielectric confinement in NC.\cite{Benchamekh14} Both are strongly affected by Si core thickness.
However, it is worth noting that the extremely long radiative decay rates presented in Fig.~\ref{fig:shellrates} suggest that in experiments, phonon-assisted recombination bypass phonon-less process.

\acknowledgments{The work of MN and AP  was financially supported by the Russian Science Foundation  grant No. 14-12-01067. KD would like to acknowledge financial support by MacGillavry Fellowship at University of Amsterdam.}

\bibliography{SiGe_QDs}

\clearpage

\section{Supplementary}\label{sec:suppl}

\begin{figure}
\includegraphics[width=9cm]{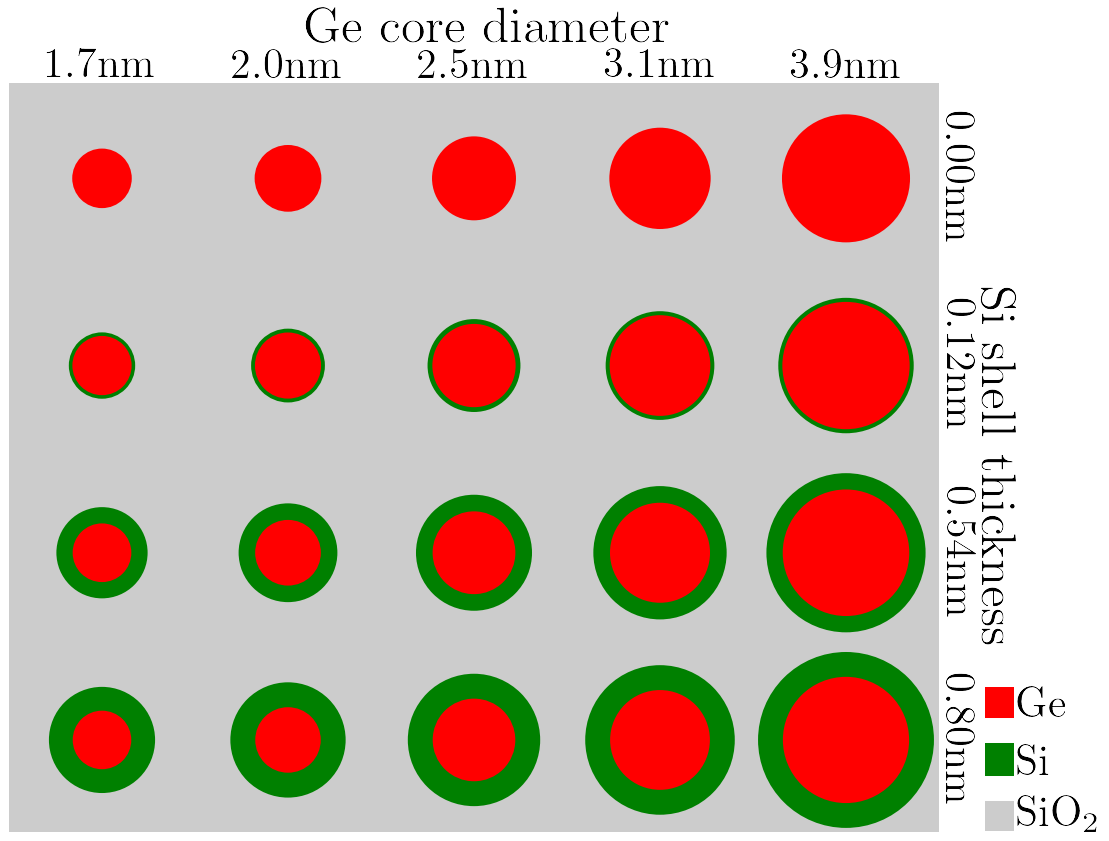}
\caption{(Color online) Schematic sketch of the simulated structures of Ge/Si core/shell NCs systems: a set of 5 Ge NC core sizes of 1.7, 2.0, 2.5, 3.1 and 3.9~nm, each core size including  Si shell thicknesses 0.00, 0.14, 0.50 and 0.80~nm. }\label{fig:samples}
\end{figure}

In supplementary material we present calculations for the core-shell NCs for 5 Ge core sizes of 1.7, 2.0, 2.5, 3.1 and 3.9~nm and for the Si shell thicknesses 0.00, 0.14, 0.50 and 0.80~nm. The systems are sketched in Fig.~\ref{fig:samples}.

\begin{figure*}
\includegraphics[width=\linewidth]{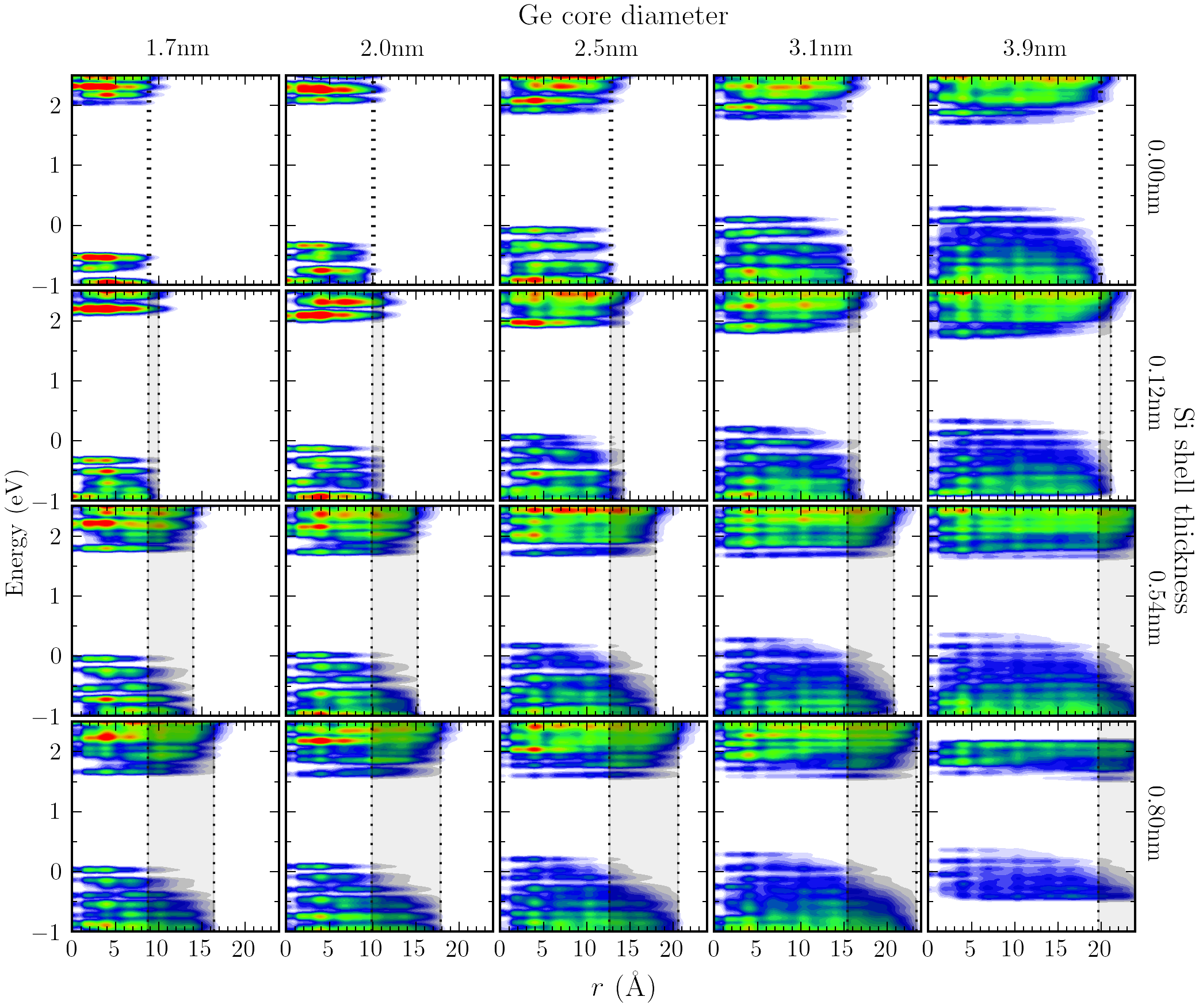}
\caption{(Color online) Real-space resolved density of states for electrons and holes. Similar to Fig.~\ref{fig:shellbandstructure}(b), but for the set of NCs with changing core radius and shell thickness}\label{fig:bandstructure_r}
\end{figure*}

Figure~\ref{fig:bandstructure_r} shows the real-space projection of the electronic density for hole and electron states. Color indicates high (red) to low (blue) DOS and the vertical gray lines denote edge of the core and shell, respectively, with the shell area in gray rectangle. 

\begin{figure*}
\includegraphics[width=\linewidth]{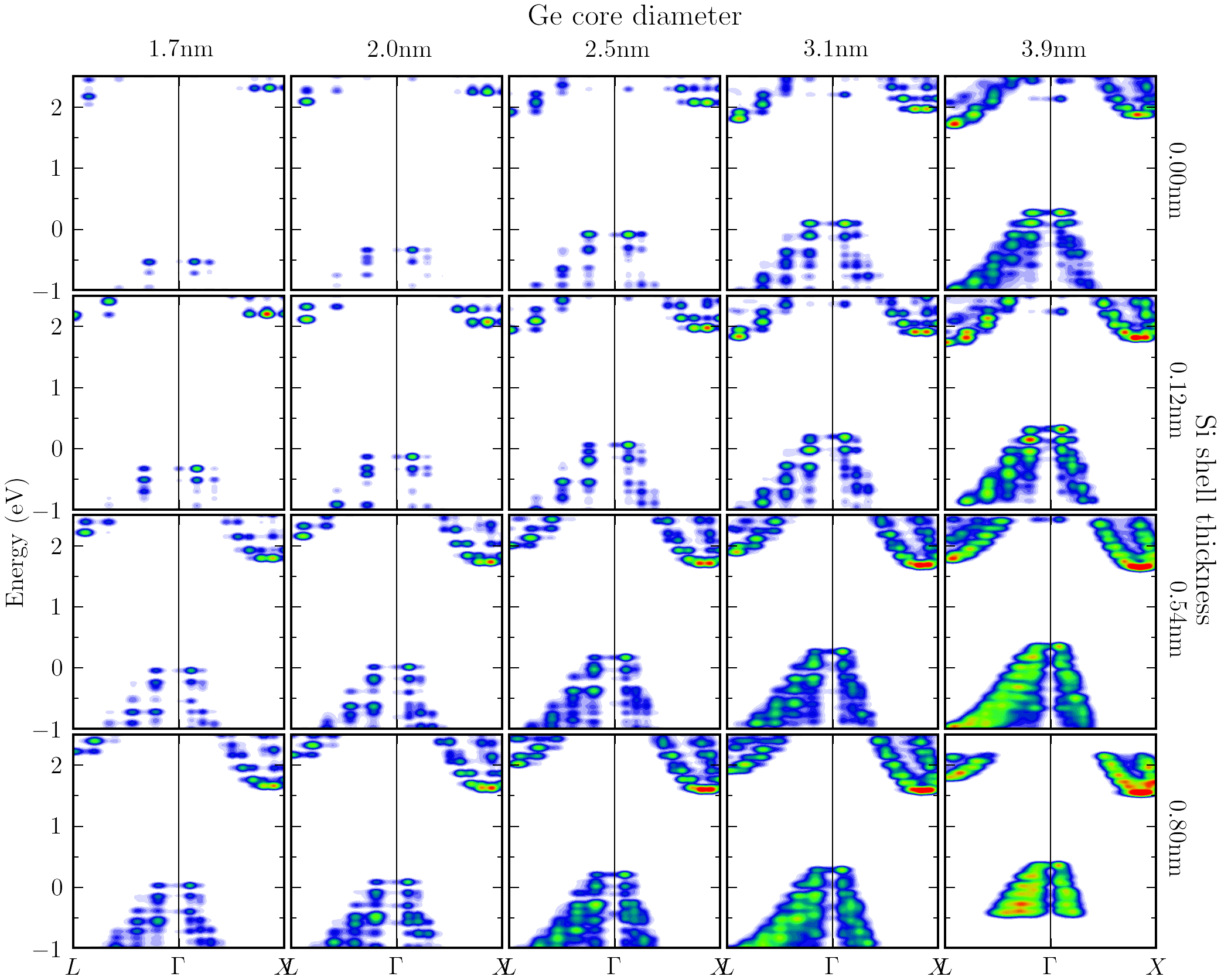}
\caption{(Color online) $k$-space resolved density of states for electrons and holes. Similar to Fig.~\ref{fig:shellbandstructure}(c), but for the set of NCs with changing core radius and shell thickness}\label{fig:bandstructure_k}
\end{figure*}

Figure~\ref{fig:bandstructure_k} shows the reciprocal space projection of the electronic density for hole and electron states. Color indicates high (red) to low (blue) DOS. 

\begin{figure*}
\includegraphics[width=\linewidth]{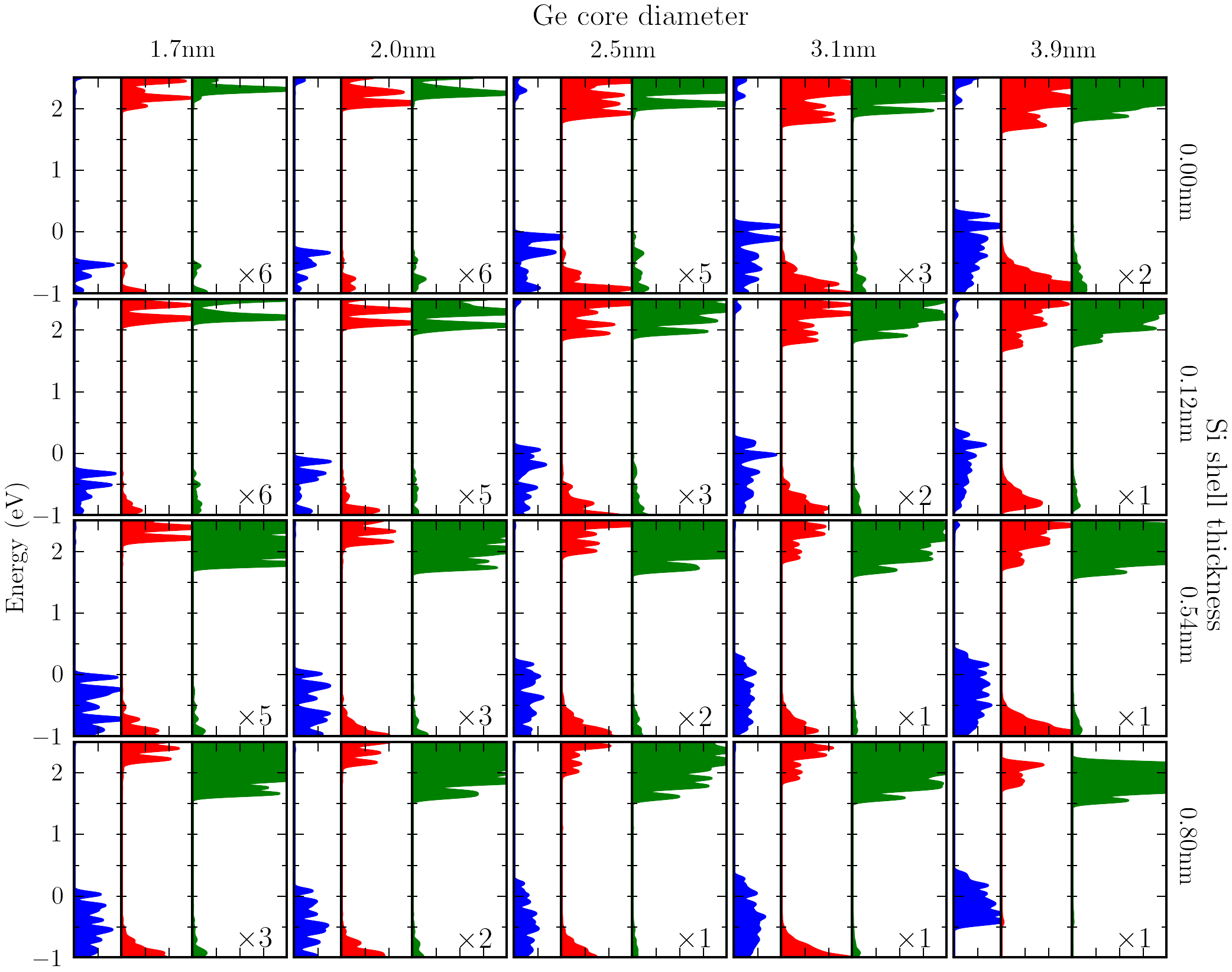}
\caption{(Color online) Valley-resolved DOS as a function of both Ge core diameter and Si shell thickness. Similar to Fig.~\ref{fig:shellbandstructure}(d), but for the set of NCs with changing core radius and shell thickness. To show the whole range of the DOS in all plots, we multiply the calculated DOS to the factor which is indicated in the right bottom corner of each panel.}\label{fig:levels}
\end{figure*}

In Fig.~\ref{fig:levels} we show the quantitative valley compositions obtained by integrating the reciprocals-space carrier density in the regions near $\Gamma$, $X$ and $L$ points. To show the whole range of the DOS in all plots, we multiply the calculated DOS to the factor which is indicated in the right bottom corner of each panel.

\begin{figure*}
\includegraphics{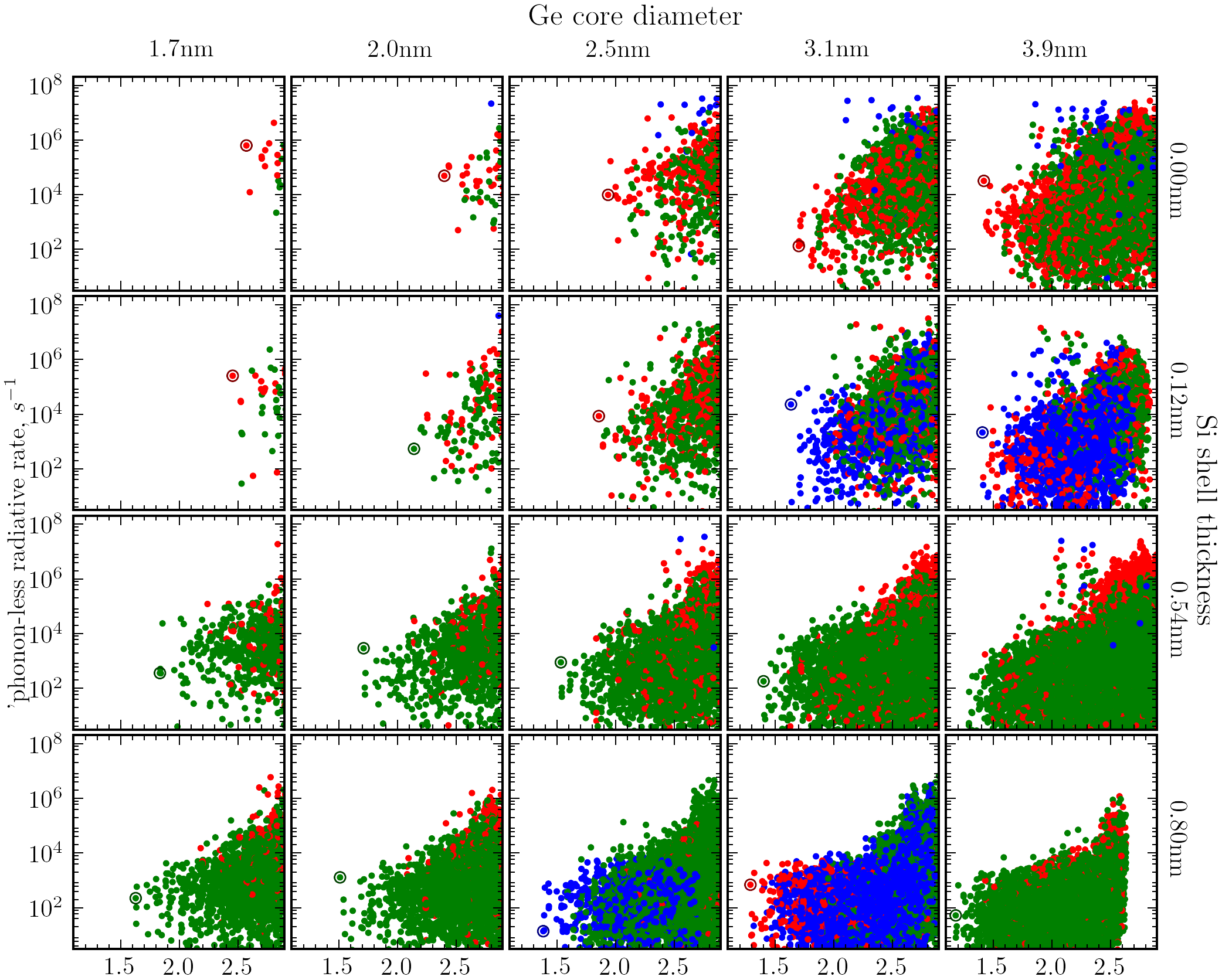}
\caption{(Color online) Radiative phonon-less transition rates resolved in k-space. Similar to the Fig.~\ref{fig:shellrates}(a).}\label{fig:ratesk}
\end{figure*}

The calculated phonon-less radiative transition rates are presented in Fig.~\ref{fig:ratesk}. Each transition is shown by a colored dot. In order to visualize the valley structure of the electron states, the color shows the amplitude (over 40\% component) in the corresponding valleys, i.e. the blue component corresponds to the electron density from over 40\% in $\Gamma$ point, green component to $X$ point and red to $L$ point. 

\begin{figure*}
\includegraphics[width=\linewidth]{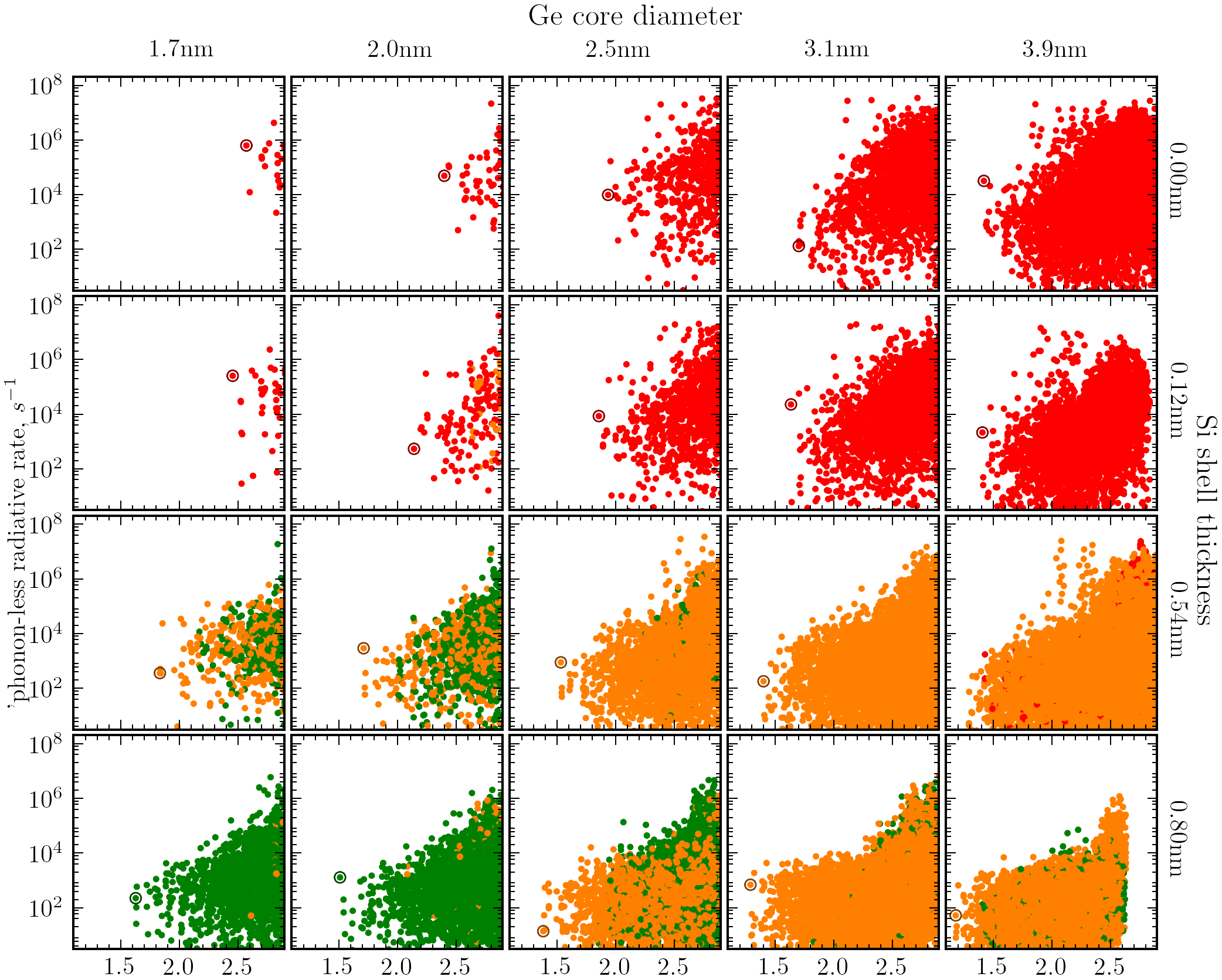}
\caption{(Color online) Radiative phonon-less transition rates resolved in color in real-space. Similar to the Fig.~\ref{fig:shellrates}(b).}\label{fig:ratesr}
\end{figure*}

In Fig.~\ref{fig:ratesr} we show the real-space-colored transition rates with color signifying more than 40\% localization of the electron density component in Ge core (red), Si shell (green) or delocalized equally in between (orange). 

\clearpage
\section{Tight-binding parameters}
In Table~\ref{tbl:TB_par} we present tight-binding parameters used in calculations. 

 \begin{table}\caption{Tight-binding parameters of Si, Ge and ``SiO$_2$'' used in calculations. Si and Ge parameters are taken from Ref.~\onlinecite{Niquet09} and SiO$_2$ are fit to reproduce bandstructure of $\alpha$-quartz. }
\label{tbl:TB_par}
 \begin{tabular*}{\columnwidth}{@{\extracolsep{\fill}}lrrr}
\hline
                 &         Si &         Ge &           ``SiO$_2$'' \\
\hline                                                                        
$              a$ & $    5.4300$ & $    5.6500$  &          \\
$         E_{s}$ & $   -2.5525$ & $   -3.4025$  & $    1.3859$  \\
$       E_{s^*}$ & $   23.4461$ & $   23.8817$  & $   27.4887$  \\
$         E_{p}$ & $    4.4859$ & $    5.3147$  & $    1.3391$  \\
$         E_{d}$ & $   14.0105$ & $   12.8753$  & $   15.1568$  \\
\hline                                                           
$       ss\sigma$ & $   -1.8660$ & $   -1.4909$  & $   -2.7825$  \\
$   s^*s^*\sigma$ & $   -4.5133$ & $   -4.8612$  & $   -5.6494$  \\
$ ss^*\sigma$ & $   -1.3911$ & $   -1.5948$  & $   -1.7632$  \\
$   sp\sigma$ & $    2.9107$ & $    2.9128$  & $    3.4027$  \\
$ s^*p\sigma$ & $    3.0682$ & $    2.9204$  & $    3.1798$  \\
$   sd\sigma$ & $   -2.2399$ & $   -2.1011$  & $   -4.6536$  \\
$ s^*d\sigma$ & $   -0.7771$ & $   -0.2356$  & $   -1.5761$  \\
$       pp\sigma$ & $    4.0848$ & $    4.3662$  & $    5.1413$  \\
$          pp\pi$ & $   -1.4921$ & $   -1.5831$  & $   -0.5472$  \\
$   pd\sigma$ & $   -1.6666$ & $   -1.6011$  & $   -2.3762$  \\
$      pd\pi$ & $    2.3994$ & $    2.3698$  & $    1.2601$  \\
$       dd\sigma$ & $   -1.8295$ & $   -1.1548$  & $   -2.7520$  \\
$          dd\pi$ & $    3.0818$ & $    2.3004$  & $    1.1614$  \\
$       dd\delta$ & $   -1.5668$ & $   -1.1939$  & $   -0.6707$  \\
\hline                                                           
$       \Delta/3$ & $    0.0185$ & $    0.1274$  & $    0.0000$  \\
\end{tabular*}
\end{table}
 \begin{table}\caption{Tight-binding of SiGe used in calculations. Parameters are taken from Ref.~\onlinecite{Niquet09}. For SiGe diagonal energies and spin-orbit constants are the same as for Si and Ge, we present only transfer matrix elements here.}
\label{tbl:TB_par2}
 \begin{tabular*}{\columnwidth}{@{\extracolsep{\fill}}lrrr}
\hline
         \multicolumn{4}{c}{Si(1)Ge(2)} \\
\hline                                                                        
$       ss\sigma$ & $      -1.6765$ & &   \\
$   s^*s^*\sigma$ & $      -4.6335$ & &   \\
$       pp\sigma$ & $       4.2193$ & &   \\
$          pp\pi$ & $      -1.5467$ & &   \\
$       dd\sigma$ & $      -1.4195$ & &   \\
$          dd\pi$ & $       2.6254$ & &   \\
$       dd\delta$ & $      -1.3938$ & &   \\
$ s_1s^*_2\sigma$ & $      -1.5094$ &  $ s_2s^*_1\sigma$ & $      -1.5031$    \\
$   s_2p_1\sigma$ & $       3.0103$ & $   s_1p_2\sigma$ & $       2.8289$    \\
$ s^*_2p_1\sigma$ & $       2.7930$ & $ s^*_1p_2\sigma$ & $       3.0630$    \\
$   s_2d_1\sigma$ & $      -2.0474$ & $   s_2d_1\sigma$ & $      -2.1399$    \\
$ s^*_2d_1\sigma$ & $      -0.5123$ & $ s^*_2d_1\sigma$ & $      -0.4639$    \\
$   p_2d_1\sigma$ & $      -1.6132$ & $   p_1d_2\sigma$ & $      -1.4341$    \\
$      p_2d_1\pi$ & $       2.4355$ & $      p_2d_1\pi$ & $       2.5711$    \\
\hline                                                                        
\end{tabular*}
\end{table}

For the strain we use parameters from Table~\ref{tbl:TB_spar}. We use simplified version of approach from Ref.~\onlinecite{Niquet09}, instead parameters $\beta$ we use parameters $\lambda$ explained in Refs.~\cite{Raouafi16,Nestoklon16}

 \begin{table}\caption{Tight-binding strain parameters used in calculations. Si, Ge and SiGe parameters are taken from \cite{Niquet09}, but we used simplified version of strain treatment and parameters $\beta$ replaced with $\lambda$ which have similar meaning. Tight-binding parameters of ``SiO$_2$'' are not renormalized by the strain.}
\label{tbl:TB_spar}
 \begin{tabular*}{\columnwidth}{@{\extracolsep{\fill}}lrrr}
\hline
                 &         Si &         Ge &                SiGe \\
\hline
$     n_{ss\sigma} $ & $    3.5670$ & $    3.5744$ &  $    3.9017$ \\
$     n_{ss*\sigma}$ & $    1.5197$ & $    1.0363$ &  $    1.0380$ \\
$     n_{sp\sigma}$  & $    2.0353$ & $    2.8820$ &  $    2.3728$ \\
$     n_{sd\sigma}$  & $    2.1481$ & $    1.8928$ &  $    1.9954$ \\
$     n_{s*s*\sigma}$& $    0.6440$ & $    1.0794$ &  $    0.8599$ \\
$     n_{s*p\sigma}$ & $    1.4665$ & $    2.6481$ &  $    1.9414$ \\
$     n_{s*d\sigma}$ & $    1.7967$ & $    2.3342$ &  $    2.0105$ \\
$     n_{pp\sigma}$  & $    2.0191$ & $    2.4058$ &  $    2.3500$ \\
$     n_{pp\pi}$     & $    2.8728$ & $    2.9503$ &  $    3.0815$ \\
$     n_{pd\sigma}$  & $    1.0045$ & $    0.5132$ &  $    0.7555$ \\
$     n_{pd\pi}$     & $    1.7803$ & $    1.6242$ &  $    1.6703$ \\
$     n_{dd\sigma} $ & $    1.7387$ & $    1.6841$ &  $    1.6698$ \\
$     n_{dd\pi}    $ & $    1.8044$ & $    2.6495$ &  $    2.2497$ \\
$     n_{dd\delta} $ & $    2.5469$ & $    3.8322$ &  $    3.0631$ \\
\hline
$       \alpha_s$ & $   -0.1336$ & $   -0.3325$ &  \\
$       \alpha_p$ & $   -0.1895$ & $   -0.4382$ &  \\
$       \alpha_d$ & $   -0.8905$ & $   -0.9049$ &  \\
$    \alpha_{s*}$ & $   -0.2437$ & $   -0.5206$ &  \\
\hline                                               %
$    \lambda_{1p}$ & $    0.1389$ & $   -0.3103$ & \\
$    \lambda_{2p}$ & $    0.0000$ & $    0.4800$ & \\
$    \lambda_{1d}$ & $    0.4394$ & $    0.2434$ & \\
\hline
 \end{tabular*}

 \end{table}

\end{document}